\documentclass[useAMS,usenatbib,usegraphicx]{mn2e}
\newcommand{\uv}{\mbox{$u$-$v$}}
\newcommand{\ex}[1]{\mbox{$\times 10^{#1}$}}

\newcommand{\Msol}{\mbox{$M$\raisebox{-.6ex}{$\odot$}}}
\newcommand{\kms}{\mbox{km s$^{-1}$}}
\newcommand{\Jb}{\mbox{Jy bm$^{-1}$}}

\newcommand{\Ra}[4]{\mbox{${#1}^{\rm h} \; {#2}^{\rm m} \; {#3}\fs{#4} $}}
\newcommand{\dec}[4]{\mbox{${#1}\degr \; {#2}\arcmin \; {#3}\farcs{#4} $}}

\newcommand{\lesssim}{\mbox{\raisebox{-0.3em}{$\stackrel{\textstyle <}{\sim}$}}}
\newcommand{\gtrsim}{\mbox{\raisebox{-0.3em}{$\stackrel{\textstyle >}{\sim}$}}}

\newcommand{\objectname}[1]{\mbox{#1}}
\newcommand\ion[2]{#1$\;${\small\rmfamily{#2}}\relax}%
\newcommand{\arcdeg}[1]{\degr}

\newcommand{\SNRX}{\mbox{SNR 4449-1}}
\newcommand{\SNRM}{\mbox{41.95+575}}

\title[Young Supernova Remnant in NGC 4449]{The First VLBI Image of the Young, Oxygen-Rich \\
Supernova Remnant in NGC 4449}

\author[Bietenholz et al.]{M.\ F.\ Bietenholz$^{1,2}$, N.\ Bartel$^1$, D.\ Milisavljevic$^3$, R.\ A.\ Fesen$^3$, 
         \and P.\ Challis$^4$, and  R.\ P.\ Kirshner$^4$ \\
$^1$Department of Physics and Astronomy, York University, Toronto, M3J 1P3, Ontario, Canada \\
$^2$Hartebeesthoek Radio Observatory, PO Box 443, Krugersdrop, 1740, South Africa \\
$^3$6127 Wilder Lab, Department of Physics \& Astronomy, Dartmouth College, Hanover, NH 03755\\
$^4$Harvard-Smithsonian Center for Astrophysics, 60 Garden Street, Cambridge, MA 02138, USA\\
}

\begin{document}


\pagerange{\pageref{firstpage}--\pageref{lastpage}} \pubyear{2010}
 
\maketitle

\label{firstpage}

\begin{abstract}

We report on sensitive 1.4-GHz VLBI radio observations of the
unusually luminous supernova remnant \SNRX\ in the galaxy NGC~4449,
which gave us the first well-resolved image of this object.  The
remnant's radio morphology consists of two approximately parallel
bright ridges, suggesting similarities to the barrel shape seen for
many older Galactic supernova remnants or possibly to SN~1987A\@.  The
angular extent of the remnant is $65 \times 40$~mas, corresponding to
$(3.7 \times 2.3) \cdot 10^{18} \, (D/3.8 \, {\rm Mpc})$~cm.  We also
present a new, high signal-to-noise optical spectrum.  By comparing
the remnant's linear size to the maximum velocities measured from
optical lines, as well as using constraints from historical images, we
conclude that the supernova explosion occurred between $\sim$1905 and
1961, likely around 1940\@.  The age of the remnant is therefore
likely $\sim$70~yr.  We find that \SNRX's shock wave is likely still
interacting with the circumstellar rather than interstellar medium.

\end{abstract}

\begin{keywords}
ISM: supernovae, supernova remnants
\end{keywords}

\section{Introduction}

In the nearby Magellanic irregular galaxy, NGC~4449, there is a
remarkable object, discovered by \citet{SeaquistB1978} in the radio.
The non-thermal radio spectrum and observations of both broad and
narrow optical lines \citep[e.g.,][]{BalickH1978, KirshnerB1980}
identified the source, \SNRX\ (also known as 1AXG~J122810+4406), as a
young supernova remnant (SNR). Further work in the optical, UV, and
X-ray confirmed both the unusual luminosity of the object and its 
SNR nature \citep{KirshnerB1980,BlairKW1983,Blair+1984}.  \SNRX\ is
quite luminous compared to known supernova remnants: at epoch 2000 its
1.4~GHz radio spectral luminosity \citep{LaceyGM2007}
and the X-ray luminosity \citep{PatnaudeF2003, Summers+2003} were
$\sim$6 and $\sim$10 times larger, respectively, than those of the
$\sim$330~yr old Cas~A, which is the most luminous Galactic remnant.
However, in the radio as well as in the X-ray, the flux density is
declining fairly rapidly, both having decreased by $\sim$50\% over the
last two decades.

Optically, \SNRX\ has been observed with a variety of ground-based telescopes
as well as with the Hubble Space Telescope (\emph{HST}),
with recent results being presented in \citet{MilisavljevicF2008}.  It
is situated in the northern outskirts of NGC~4449, near two prominent
\ion{H}{II} regions.  Its ejecta are mainly oxygen-rich and show broad
emission lines of forbidden oxygen along with both narrow \ion{H}{II}
region-like lines and broader ($v = 500$~\kms) H$\alpha$ and
\ion{N}{II} lines.  The highest velocities are seen in the forbidden
lines, whose line-widths \citet{MilisavljevicF2008} show to be
$\sim$6000~\kms.  The supernova type is not known, although various
arguments suggest a fairly massive progenitor, with a likely mass of
$\sim$25~\Msol\ \citep[see][and references
therein]{MilisavljevicF2008}, implying that it was a core-collapse
supernova (Type II or Ib,c).

The distance, $D$, to NGC~4449 is $3.8\pm0.3$~Mpc
\citep{Annibali+2008}, and we will indicate scaling with $D$ where
appropriate.
The unmixed oxygen debris, high velocity, and small size all point to
quite a young remnant, but neither the size nor the age of the remnant
has so far been well determined.  Earlier very-long-baseline
interferometry (VLBI) radio observations by \citet{deBruyn1983}
resulted only in upper limits on the size, and it was only marginally
resolved by the \emph{HST} \citep{BlairF1998, MilisavljevicF2008}.
Various estimates of the age have been published, from a young age
range of $50 \sim 100$~yr \citep{MilisavljevicF2008} to a rather older
age of $\sim$270~yr, similar to that of Cas~A \citep{Summers+2003}.

We have undertaken sensitive new VLBI observations
using a global array in order to determine the size of the remnant and
better address the questions of the remnant's age and nature.

\section{Observations and Data Reduction}
\subsection{VLBI Radio Observations}
\label{svlbi}

We observed \SNRX\ on 2008 July 19, at 1.4~GHz, using the ``High
Sensitivity Array'' which consisted of nine stations of the
NRAO\footnote{The National Radio Astronomy Observatory, NRAO, is a
facility of the National Science Foundation operated under cooperative
agreement by Associated Universities, Inc.}
Very Long Baseline Array (each 25~m diameter)\footnote{The antenna at
Fort Davis, Texas did not take part in this project}, and the Robert
C. Byrd ($\sim$105~m diameter) as well as the Effelsberg (100~m
diameter)\footnote{The telescope at Effelsberg is operated by the
Max-Planck-Institut f\"{u}r Radioastronomie in Bonn, Germany.}
telescopes.  The observations lasted 12~h, and we recorded both senses
of circular polarisation. We used 2-bit sampling at a bit rate of
512~Mbit~s$^{-1}$, for an effective bandwidth per polarisation of
64~MHz.

The data were correlated with NRAO's VLBA processor, and the analysis
was carried out with NRAO's Astronomical Image Processing System
(AIPS)\@.  The flux density calibration was done through measurements
of the system temperature at each telescope, and the antenna amplitude
gains were subsequently improved through self-calibration of the
reference sources.  Our final data are phase-referenced to the quasar
\objectname{FIRST J122657.9+434058}, which we will refer to as
J1226+4340, and whose position of RA = \Ra{12}{26}{57}{9044980}, and
dec.\ = \dec{43}{40}{58}{443280} (J2000) we take from the fifth VLBA
calibrator survey \citep{Kovalev+2007}.

\subsection{Optical Spectroscopy}

In addition to the VLBI observations, a low-dispersion optical
spectrum of \SNRX\ was obtained with the 6.5~m MMT telescope at Mount
Hopkins in Arizona on 2008 April 1 using the Blue Channel Spectrograph
\citep{SchmidtWF1989} with a 1\arcsec $\times$ 150\arcsec\ slit and a
300 lines mm$^{-1}$ 4800~\AA\ blaze grating.  The slit was placed at
the parallatic angle.  The spectrum of the SNR was extracted using
optimal extraction techniques covering a 2\arcsec\ region centered on
the SNR and spanned the wavelength range from 3500~\AA\ to 8000~\AA\
with a FWHM resolution of 7~\AA.  Data were reduced and calibrated
employing standard techniques in IRAF and standard stars from
\citet{Strom1977}. No second order filter was used, so some second
order overlap at the red end of the spectrum is possible. However,
careful cross-calibration with standard stars of different colours has
minimised problems this might cause in the flux calibration
\citep[see][]{Matheson+2008x}.

\section{Results}
\subsection{VLBI Image}
\label{svlbir}

\begin{figure*}
\centering
\includegraphics[width=0.7\textwidth]{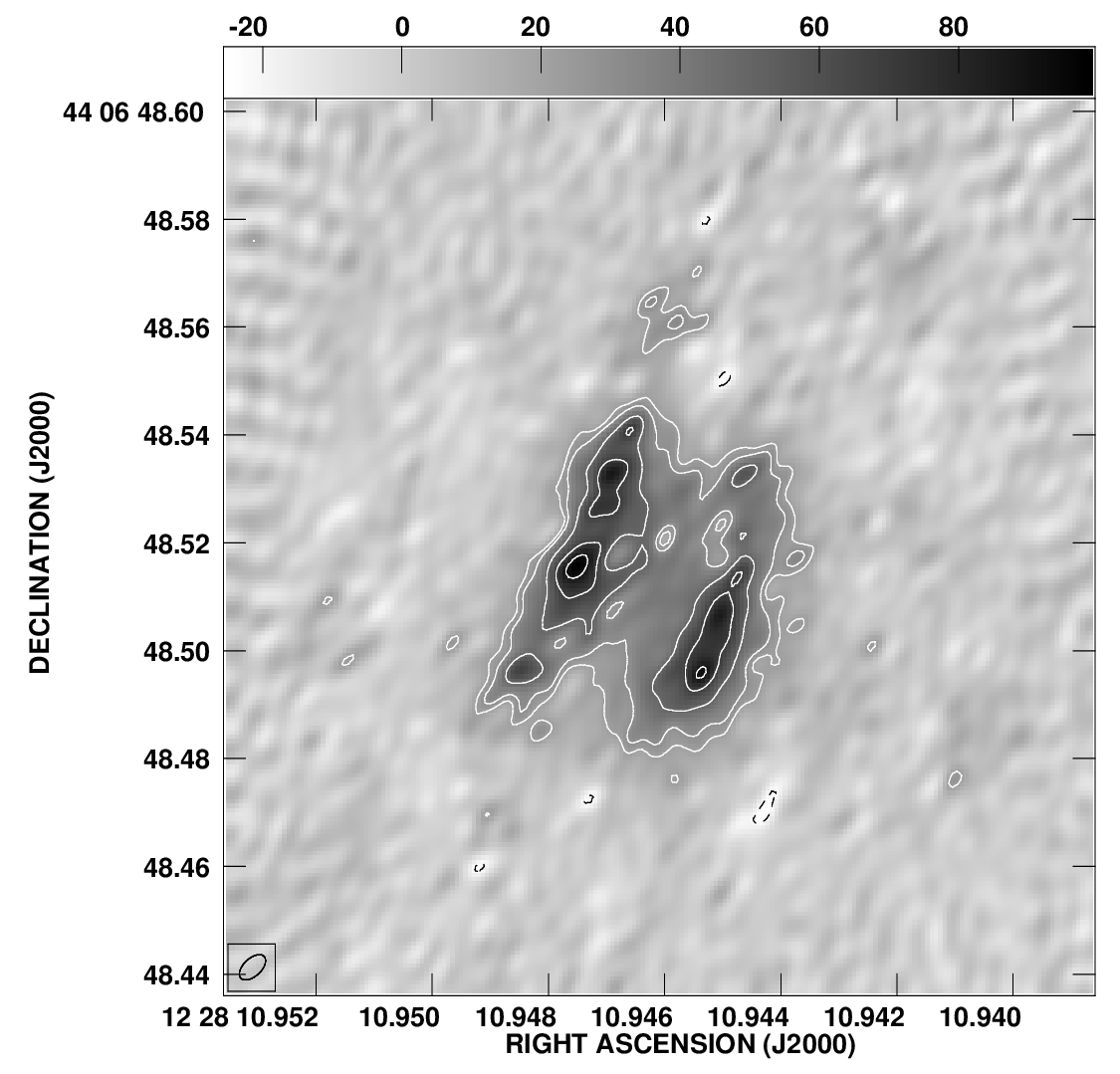}
\caption{A 1.44~GHz VLBI image of \SNRX\@.  The contours are drawn at
$-22$, 22, 30, 50, 70 and 90\% of the peak brightness, which was
98~$\mu$\Jb.  The lowest contour is three times the off-source
brightness rms, which was 7.1~$\mu$\Jb, although we estimate that the
effective on-source rms is likely $\sim$50\% higher.  The FWHM of the
convolving beam was $5.9 \times 3.3$~mas at $-47$\degr, and is
indicated at lower left.  The image was deconvolved using
multi-resolution CLEAN.}
\label{fsnrx}
\end{figure*}

We show a VLBI image of \SNRX\ in Fig.~\ref{fsnrx}.  This image was
obtained using natural weighting for the highest signal-to-noise
ratio, and a multi-scale CLEAN deconvolution \citep{WakkerS1988}.  We
discuss the deconvolution process and the uncertainties inherent
therein in more detail in Section~\ref{sdeconv} below.  The total flux
density determined from the VLBI observations was
$6.5_{-0.7}^{+1.4}$~mJy, where we have estimated the uncertainty in
the VLBI flux density scale as 10\%, and allowed an additional 20\%
upwards uncertainty in the total flux density as it is possible that
we did not recover all the extended emission in our VLBI image.  The
off-source rms background brightness was 7~$\mu$\Jb, and the peak/rms
dynamic range was $\sim$14\@.  The FWHM of the elliptical-Gaussian
convolving beam was $5.9 \times 3.3$~mas at $-47$\degr.

\SNRX\ shows a two-part radio morphology, with two bright parallel
ridges of emission.  Their length is $40 \sim 60$~mas, and their
peak-to-peak separation is $\sim$30~mas, corresponding to 2.8\ex{18}
($D$/3.8~Mpc)~cm and 2.0\ex{18} ($D/3.8$~Mpc)~cm respectively.  The
morphology might be interpreted as an elliptical shape with a deficit
of brightness near the ends of the major axis, a major-axis angular
diameter $\sim$65~mas at p.a.\ $\sim$155\arcdeg, and a minor-axis one
of $\sim$30~mas. Note that the minor axis is better determined than
the major one, which might be somewhat larger than 65~mas.  This
angular extent corresponds to $(3.7 \times 1.7) \cdot 10^{18} \, (D$/3.8
Mpc)~cm.

Our angular diameter estimate is slightly larger than the 28 - 37 mas
estimates derived from optical \emph{HST} measurements, where the
remnant was only marginally resolved given the instrumental FWHM of 42
mas \citep{BlairF1998,MilisavljevicF2008}.

\subsection{Imaging and Deconvolution Errors}
\label{sdeconv}

The process of deconvolution can lead to errors in the image which are
larger than the noise.  As the signal-to-noise ratio in our image is
rather low we carried out some additional tests to verify the
reliability of our image.

To determine the effect of the choice of deconvolution method on the
image, we tried various methods.  Over the deconvolved region the rms
brightness difference between a multi-scale and single-scale CLEAN
image was 8~$\mu$\Jb, and that between a maximum entropy (AIPS task
VTESS) and the multi-scale clean image was 10~$\mu$\Jb.  These
differences suggest that the deconvolution errors are between 1 and
1.5 times the rms background brightness.

One might expect a young supernova remnant to show an edge-brightened
and approximately circular structure in the radio.  Could \SNRX\ be
circularly symmetric despite the appearance of the image?  In
particular, it has been pointed out that some apparent deviations from
circular symmetry can be due to deconvolution errors, which are
coupled to the \uv~plane sampling \citep{Heywood+2009}, and which tend
to be bilaterally symmetric, much like the structure observed in
\SNRX\@.  Could this be the origin of the observed bilateral
structure? To test this possibility, we produced model visibilities
from a simple, circularly symmetric shell model of an SNR chosen to
approximately match \SNRX\ in size and location, with a total flux
density of 6~mJy and random noise added at a level comparable to that
in our observed visibilities.  The specific model used was the
projection of a three-dimensional shell of uniform emissivity, with an
outer angular radius of 28.2~mas, and an inner radius being 80\% of
the outer one.  Such a model was found to be appropriate for the
supernova SN~1993J in M81 \citep{SN93J-3}.  When deconvolved, these
model data produced a very circular image with a fairly uniform
brightness around the rim, suggesting that the distortions seen in the
image of \SNRX\ are not due to deconvolution errors or gaps in
\uv~coverage, but are intrinsic to the source.  In summary, our tests
confirm that the source structure in our VLBI image is reliable, and
suggest that if \SNRX\ has a shell morphology, it must be fairly
distorted.

\subsection{Optical Spectrum}
\label{sspect}

\begin{figure*}
\centering
\includegraphics[width=0.60\linewidth]{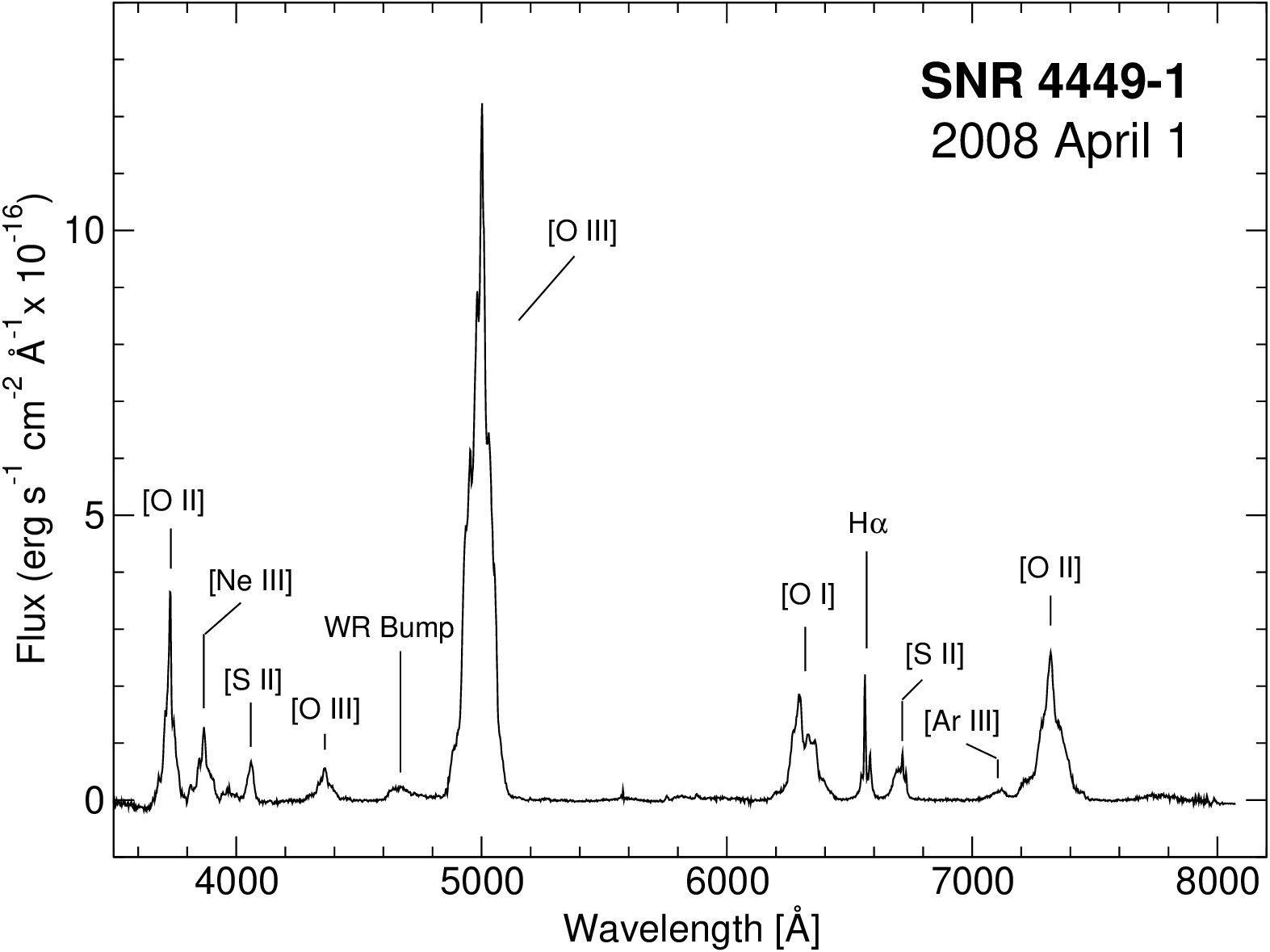} 
\includegraphics[width=0.60\linewidth]{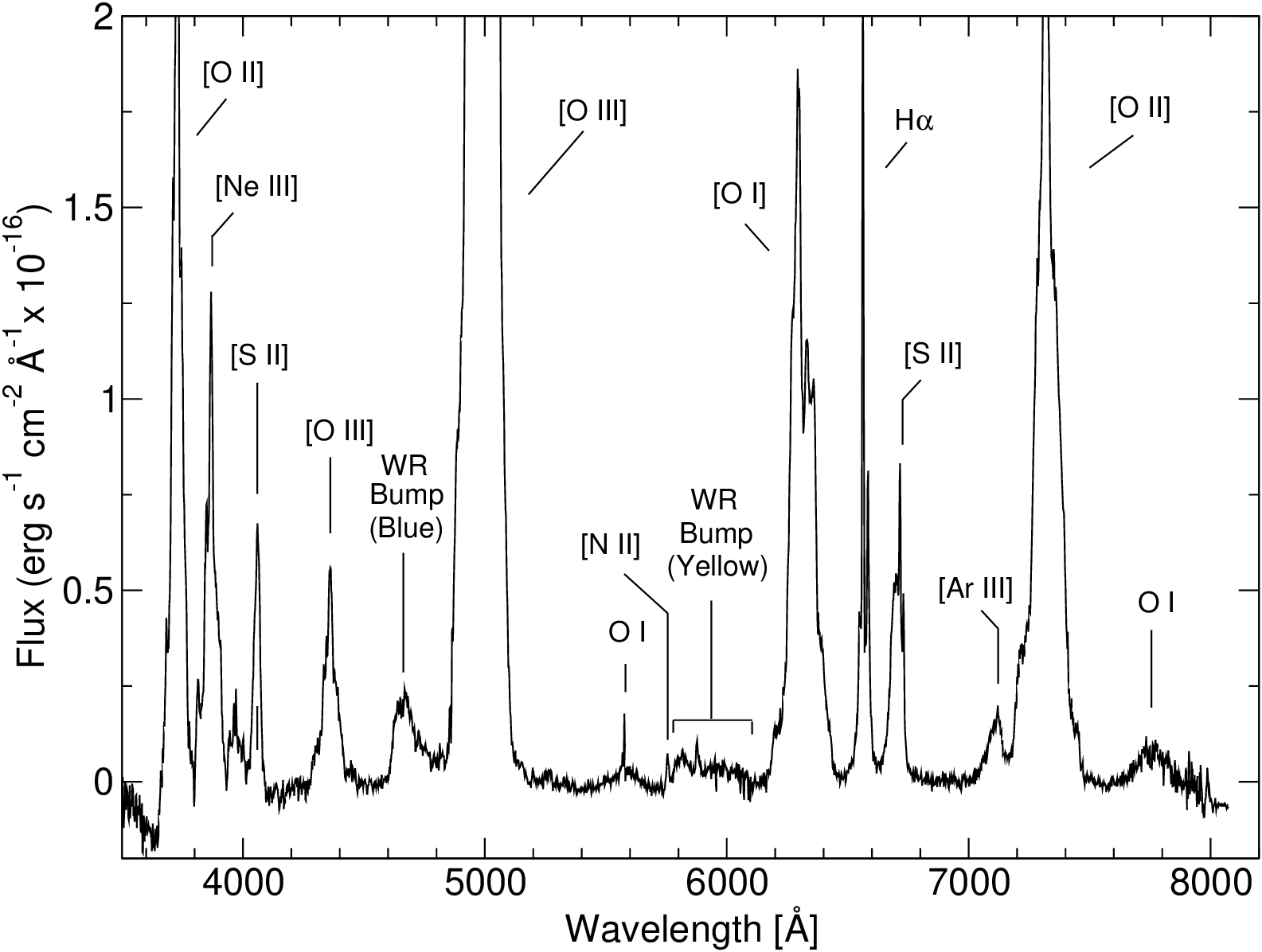}
\caption{April 2008 optical spectrum of \SNRX. The entire spectrum is shown in the top panel,
with a magnified plot presented in the bottom panel to
show the remnant's weaker emission features. }
\label{fspix}
\end{figure*}

In Fig.~\ref{fspix}, we show our April 2008 optical spectrum of \SNRX,
with the listed wavelengths given in the rest frame of the supernova
as measured by H$\alpha$ emission observed at 6567~\AA\ (velocity =
190~\kms) originating from a neighbouring \ion{H}{II} region
\citep[see Fig.\ 2 in][]{MilisavljevicF2008}. A blue continuum likely
associated with a bright star cluster clearly visible in \emph{HST}
images \citep{MilisavljevicF2008} has been subtracted.

Dominating the spectrum of \SNRX\ are broad [\ion{O}{I}]
$\lambda\lambda$6300,6364, [\ion{O}{II}] $\lambda\lambda$7319,7330,
[\ion{O}{III}] $\lambda\lambda$4959,5007 emission lines consistent
with the O-rich ejecta of a young SNR\@. Less prominent and not as
broad are emission lines from [\ion{S}{II}] $\lambda\lambda$4069,4076
and $\lambda\lambda$6716,6731, [\ion{Ne}{III}] $\lambda$3869, and
[\ion{Ar}{III}] $\lambda$7136. Contributions from nearby \ion{H}{II}
regions are also observed as narrow lines of H$\alpha$, [\ion{N}{II}]
$\lambda\lambda$6548,6583, and [\ion{S}{II}]
$\lambda\lambda$6716,6731.

In most regards our spectrum shows no significant changes in the
emission lines from the 2002 and 2007 spectra presented in
\citet{MilisavljevicF2008}. However, our broader wavelength coverage
and improved signal-to-noise ratio reveals some previously unreported
features in the remnant's optical emission.  For example, faint and
broad emission centred around 7770~\AA\ is now seen. This emission's
velocity width and line profile shape match those of the forbidden
oxygen lines and it is thus likely associated with \ion{O}{I}
$\lambda$7774\@.  Faint and broad emission is also observed around a
narrowly peaked \ion{O}{I} $\lambda$5577 line. While the narrow
emission may in part be an artifact of poor night sky subtraction, the
underlying broad emission is likely real and associated with the
remnant.

The increased sensitivity of the MMT spectra also allows for improved
estimates of expansion velocities of the optically emitting material.
Measured from the half width at zero intensity along the red wing of
the [\ion{O}{III}] $\lambda$5007 line, we estimate an expansion
velocity of $\approx 6500$~\kms.  This velocity is marginally larger
than the 6000 km s$^{-1}$ value reported by
\citet{MilisavljevicF2008}.  In addition, the broad emission centred
around H$\alpha$ can now be seen to extend from 6525 to 6605~\AA,
which is 20~\AA\ larger than that cited by \citet{MilisavljevicF2008}
and implies expansion velocities up to $\sim$1000 km s$^{-1}$ for the
H$\alpha$ and [\ion{N}{II}] $\lambda\lambda$6548,6583 lines.

A relatively strong and broad emission feature, the so-called ``blue
WR bump'' centred between $4600 - 4700$~\AA\ and associated with the
\ion{N}{III} $\lambda\lambda$4634,4642, \ion{N}{V}
$\lambda\lambda$4605,4622, and \ion{C}{III}/\ion{C}{IV}
$\lambda\lambda$4650,4658 lines, can also be seen in the spectrum,
indicating the presence of Wolf-Rayet (WR) stars of the WN type.  This
WR emission feature was previously seen in both ground-based and {\sl
HST} spectra and is likely from a WR population located within the OB
star cluster coincident with the SNR \citep{MilisavljevicF2008}.

Broad emission spanning $5740-6100$ \AA\ is also present with an
unidentified peak around 5816~\AA. The overall position, broadness,
and strength resembles the ``yellow WR bump'' due to \ion{C}{III}
5696, \ion{C}{IV} $\lambda\lambda$5801,5812, and \ion{He}{I}
$\lambda$5876 emission lines seen in the spectra of WR stars mainly of
the WC type. The fact that this emission feature was not reported
previously may indicate that our particular long slit position
detected neighbouring WC stars not located within the OB star cluster
coincident with SNR 4449-1.

\section{Discussion}
\label{sdiscuss}

We have produced the first VLBI image of the unusually bright young
supernova remnant, \SNRX, in the galaxy NGC~4449\@.  The morphology of
the remnant does not correspond to a circularly symmetric shell as
might be expected for a young supernova or supernova remnant.
Instead, our VLBI image (Fig.~\ref{fsnrx}) shows a two-part morphology
with two parallel, almost straight ridges of emission.  The overall
emission region is somewhat elongated, extending over a region of
about $65 \times 40$~mas, corresponding to $(3.7 \times 2.3) \cdot
10^{18} (D/3.8$~Mpc)~cm (see Section~\ref{svlbir}).
We will discuss the morphology more extensively below, but first we
turn to estimating \SNRX's age, which has so far been an outstanding
question.

Based on the presence of \SNRX\ in images from 1961, the remnant must
at present (2010) be at least 49~yr old \citep{MilisavljevicF2008}\@.
An upper limit on its age, however, is much less certain, with
\citet{MilisavljevicF2008} estimating $\lesssim$100~yr
while \citet{Summers+2003} suggest it could be as old as $\sim$270~yr.

From the [\ion{O}{III}] line emission in our optical spectrum, we find
a maximum expansion velocity of 6500~\kms.  An estimate of the
remnant's age can be obtained by dividing remnant's radius, determined
from the VLBI image, by the maximum expansion velocity.  This estimate
involves a number of assumptions, whose effect on the age estimate we
discuss below.  Firstly it assumes constant velocity expansion, or in
practice, no significant deceleration, since acceleration is
physically unlikely.  Secondly, it asssumes that \SNRX's O-rich
material is (approximately) co-located with the bright radio shell
detected in the VLBI image.  Thirdly, it assumes that the remnant's
extent along the line of sight, or the $z$ direction, can be estimated
from the angular extent in the sky-plane as revealed by the VLBI
image.

In particular, we estimate a range of possible $z$-axis extents by
taking the major and minor axis angular extents from the VLBI image
coupled, respectively, with the $2\sigma$ range of distance estimates
of 4.4 and 3.2~Mpc, giving a range of $z$-axis linear diameters of
(4.3 to 1.9) $\times 10^{18}$~cm.
The velocity of 6500~\kms\ then implies a range of ages (from the
epoch of the VLBI observations of 2008.6) of 104 to 47 years, or an
explosion epoch between 1905 to 1962, with a median value of 1934\@.
We note that in view of the unusual morphology of the remnant, it is
possible that the extent along the line of sight could be either
larger or smaller than our range, resulting in either an expansion age
which could possibly be outside the range we give, but we give other
arguments which limit the range of possible explosion dates below.

This process of comparing the optical expansion velocity to the size
determined from the VLBI image is likely to \emph{overestimate} the
true age for two reasons.  First, the remnant's outer shock front,
which corresponds approximately to the outer edge of the radio
emission, is likely outside of the bulk of the ejecta, and is thus
moving more rapidly than the [\ion{O}{III}] emitting gas.  However,
the highest velocity optical emission has been convincingly shown to
have velocities close to (although slightly smaller than) that of the
forward shock for SN~1993J \citep{SN93J-4} and SN~1979C \citep{SN79C}.
Second, the object's considerable and long-lived luminosity suggests a
strong interaction with the local circumstellar medium (CSM), leading
to the conclusion that some deceleration of the shock front and ejecta
is certain to have occurred.  The most likely value of the explosion
epoch, therefore, is somewhat later than our median value of 1934
and we adopt a round value of 1940\@.

Our lower limit on the explosion epoch of 1905 is therefore
conservative.  In fact, as argued by \citet{MilisavljevicF2008}, it
was probably later than 1913, because \SNRX\ is not clearly
distinguishable on a deep, blue-sensitive plate from that year, and
would be expected to have been quite bright if it had exploded before
then.

If \SNRX\ had undergone strong deceleration, such as SN~1987A
\citep[e.g.,][]{Gaensler+1997}, it could have exploded even later than
1963\@.  However, the 1961 detection mentioned above proves that it
must in fact have exploded before 1961, and we can conclude that no
very strong deceleration has occurred.

We can therefore say that \SNRX\ most likely exploded around 1940,
almost certainly between 1905 and 1961\@.  In other words, \SNRX's age
is likely $\sim$70~yr, which is near the middle of the range estimated
by \citet{MilisavljevicF2008}.  This young age makes \SNRX\ a very
interesting case, since the youngest known Galactic SNRs are
\objectname{Cas A} with an age of $\sim$330~yr and \objectname{SNR
G001.9+00.3}, thought to be about 150~yr old \citep{Reynolds+2008,
Green+2008}\footnote{We note also \objectname{GAL~000.570-00.018}
(G0.570-0.018), whose nature is uncertain, but if interpreted as an
SNR, probably has an age of $80 \sim 160$~yr \citep{Renaud+2006}.}.

Radio emission from the remnant of a high mass progenitor should
decline with time while the ejecta are interacting with the
progenitors's dense stellar wind, whose density decreases
approximately with the inverse square of the radius. This appears to
be the case with \SNRX, whose radio lightcurve has shown a steady
decline since at least 1970 \citep{LaceyGM2007}. We find that the
observed 1.4-GHz lightcurve is consistent with a continuous decline
with time as approximately $t^{-1.7}$ (average value between 1980 and
the present, assuming an explosion date of 1940) and that the total
flux density
from our recent VLBI observations of $6.5_{-0.7}^{+1.4}$~mJy lies
almost exactly on the extrapolation of Lacey et al.'s 1.4-GHz radio
lightcurve.  The remnant's present shock radius of $\sim$1.5\ex{18}~cm
is comparable to the size of the wind bubbles around massive
stars. There is no sign of any increase in the radio lightcurve which
might suggest that the shock had encountered the interstellar medium
(ISM) where the density is independent of the radius. The progression
of the shock from the CSM to the ISM, which for typical remnants
occurs at age of order 100~yr, should result in the flux density
increasing relatively rapidly until the swept-up mass becomes roughly
equal to the ejecta mass \citep[e.g.,][]{CowsikS1984}.
We can compare the radio luminosity of \SNRX\ with that of other young
supernovae (SNe) and SNRs of similar age.  \SNRX's present 1.4-GHz
spectral luminosity is $1.1 \times 10^{26} \, (D/{\rm 3.8 \, Mpc})^2$
erg~s$^{-1}$~Hz$^{-1}$.  Only two optically-identified supernovae of
comparable age that have been detected in the radio: SN~1923A and
SN~1950B, which have spectral luminosities of $\sim$5\% and
$\sim$13\%, respectively, that of \SNRX\ \citep{EckCB2002}.
Extrapolating their 1.4-GHz lightcurves of some of the more radio
luminous SNe detected in the last few decades to $t = 70$~yr, we find
that most are considerably fainter than \SNRX\@.  Even a luminous and
long-lived example like SN~1986J has an extrapolated 1.4-GHz
luminosity only 11\% that of \SNRX\ \citep{SN86J-2}.
Of the optically detected SNe, only SN~1979C, which has shown a
relatively shallow flux density decay, has an extrapolated 1.4-GHz
luminosity larger than \SNRX, being $\sim 3\times$ that of \SNRX\
\citep{SN79C-shell}.
The most luminous example is SN~1982aa, the radio supernova in Mrk~297
(not optically detected) which was likely a type II, which has an
extrapolated 1.4-GHz luminosity of $\sim 30\times$ that of \SNRX\
\citep{Yin1994}.
In other words the radio luminosity of \SNRX\ is high compared to that
of most supernovae at age 70~yr, and even compared to of the subset of
SNe detected in the radio, but is lower than the most luminous known
examples.  The high luminosity suggests a dense CSM and therefore that
the progenitor underwent a phase of high mass-loss phase in the period
leading up to the SN explosion.

The progenitor of \SNRX\ was likely quite massive.
\citet{MilisavljevicF2008} discuss the nature of the progenitor on the
basis of the spectrum and the cluster turnoff, and conclude that
\SNRX's progenitor likely had a mass $\gtrsim 20$~\Msol.  They favour
a luminous blue variable (LBV) over a WR star because WR stars
typically do not have a CSM dense enough to reproduce the optical
spectrum of \SNRX\@.  The high and long-lived radio luminosity also
requires strong interaction with a dense CSM, therefore supporting the
LBV hypothesis.  We conclude that \SNRX\ likely exploded as a luminous
blue variable, or perhaps as a red supergiant.

There is no sign in \SNRX\ of a central radio component, such as might
be associated with the neutron-star or black-hole compact remnant of
the supernova.  For example, a young and energetic pulsar might be
expected to produce a very bright wind nebula.  Such a central
component is seen in SN~1986J, although its association with the
compact remnant is not certain \citep{SN86J-2}.  A central component
has not so far been seen in any other young supernova
\citep{NB-COSPAR}.  If we take the peak brightness in \SNRX's central
region of 53~$\mu$\Jb\ as a limit then we can say that the 1.4-GHz
spectral luminosity of any unresolved central source in \SNRX\ must be
$\lesssim 1.3 \times 10^{24}$~erg~s$^{-1}$~Hz$^{-1}$ or $\lesssim$20\%
that of the Crab Nebula.

As we have noted, the observed morphology of \SNRX\ (Fig.~\ref{fsnrx})
is somewhat different than that seen in the small sample of resolved
radio supernovae \citep[see, e.g.,][]{SNVLBI-Bologna, VLBA10th}.  How
can it be interpreted?  \SNRX's observed morphology is similar to the
barrel shape seen in some older Galactic SNRs \citep{KestevenC1987,
Gaensler1998}, which is thought to be due to interaction with the
large-scale Galactic magnetic field, with the symmetry axis being
preferentially parallel to the Galactic plane \citep{Gaensler1998}.
In the case of \SNRX, which as we have shown is interacting with the
CSM still, the influence of NGC~4449's large-scale galactic field is
likely to be small, especially given the fact that \SNRX\ is over an
order of magnitude smaller than the barrel-shaped Galactic SNRs.  We
do note, however, that \SNRX's axis of (approximate) symmetry is also
parallel to the large-scale field, which has an orientation of
$\sim$45\arcdeg\ at the location of \SNRX\ \citep[][note that
NGC~4449's magnetic field seems not to be aligned with the spiral arms
as is usually the case]{Chyzy+2000a, Chyzy+2000b}.

The morphology of \SNRX\ is perhaps also reminiscent of the morphology
seen in the radio for SN~1987A, which shows a ring-like structure with
a prominent hot-spot on either side \citep[e.g.,][]{Zanardo+2010,
Potter+2009, Tingay+2009, Ng+2008}.  In SN~1987A, this structure is
thought to be due to the radio emission arising in a thick equatorial
belt, thought to have been generated by the collision of a fast wind
produced by the progenitor in its final blue supergiant stage with the
slow wind from the preceding red supergiant stage
\citep{Chevalier1988}.  This mechanism, however, does not seem
applicable for \SNRX, which must have exploded directly after a period
of intense mass loss, as only interaction with a slow dense wind can
give rise to \SNRX's high radio luminosity which is several
orders of magnitude above that of SN~1987A\@.

Finally, we note that the two-part morphology of \SNRX\ is also
somewhat reminiscent of \SNRM, the brightest and most compact radio
source in M82 \citep[e.g.,][]{McDonald+2001b, Bartel+1987}. 
Although the unusual morphology of \SNRM\ has led some authors to
question whether it is in fact supernova-related, it seems that on the
balance the likeliest explanation is still that \SNRM\ is a supernova
in a very dense environment.  Its age is not known, but must be at
least $\sim$40~yr, with the low expansion velocity suggesting an age
of $\sim$75~yr, quite comparable to \SNRX\@.  \SNRM\ likely exploded
in an even denser environment, since its radio luminosity is about 5
times higher, its rate of flux density decay about twice, and its
expansion speed about half those, respectively, of \SNRX.  On the
basis of this evidence, it would seem reasonable to suppose that 
\SNRM\ is interacting with a particularly dense interstellar medium,
while \SNRX\ is at an earlier stage of its evolution and is still
interacting with its CSM\@.  Although difficult, VLBI observations of
higher sensitivity would be useful to clarify the morphology of \SNRX,
and should be feasible with the planned increases in bandwidth for the
VLBA \citep{Ulvestad+2010}.

\section{Summary}

We report on high-sensitivity array VLBI observations and a high
signal-to-noise optical spectrum of \SNRX\ in the galaxy NGC~4449\@.
Our VLBI image shows, for the first time, the morphology of \SNRX,
which consists of two parallel bright ridges.  The morphology suggests
a barrel-shape, as is seen in many Galactic supernova remnants, or
perhaps a ring-shape, similar to that seen in SN~1987A\@.  From the
image, we determine also the first direct measurement of the size of
the remnant, with the extent (in the sky plane) being $(3.7 \times
2.3) \cdot 10^{18} \, (D$/3.8 Mpc)~cm.  This extent, when combined with the
velocities obtained from the optical spectrum, allows us to deduce an
explosion epoch between $\sim$1905 and 1961, with a likely value near
1940, or a present age of 70~yr.  We also conclude that \SNRX\ is
probably still interacting with its circumstellar medium.

\section*{Acknowledgments}

The optical observations reported here were obtained at the MMT
Observatory, a joint facility of the Smithsonian Institution and the
University of Arizona.  Supernova studies at the Harvard College
Observatory are supported by NSF grant AST09-07903, and at York
University by NSERC.

\bibliographystyle{mn2e}

\bibliography{mybib1,snr4449-temp}

\end{document}